\documentstyle[referee]{mn}
\input epsf
\newif\ifAMStwofonts

\title{Test problems for Relativistic Magnetohydrodynamics}
\author[S.S.Komissarov]
  {S.S.Komissarov, \\
Department of Applied Mathematics, 
The University of Leeds, 
Leeds LS2 9JT 
}

\begin{document}
\label{firstpage}
\maketitle

Growing interest to numerical relativistic MHD has led to a number 
of new computer codes being developed by various research groups 
around the globe. A number of demanding test problems for such codes
have been described in Komissarov(1999). Unfortunately, some of the 
parameters given in Table 1 of this paper are incorrect. I would like 
to apologies to my colleagues and thank 
Charles Gammie from the University of Illinois for noticing this most 
embarassing confusion and letting me know about it.  
Here is the updated table with the modified parameters indicated 
by the exclamation symbol.

\begin{table*} 
   \begin{center} 
   \begin{tabular}{|c|l|l|l|l|} 
   \hline 
   \hline
   \ Problem &  Left state &  Right state & Grid & Time \\ 
   \hline 
   \hline
    Fast Shock 
                 & $ u^i = (25.0,0.0,0.0)$ 
                 & $ u^i = (1.091,0.3923,0.00)$  
                 & $ n = 40 $ & $ t = 2.5 $      \\ 
   $(\mu = 0.2)$ 
                 & $ B^i = (20.0,25.02,0.0) $ 
                 & $ B^i = (20.0,49.0,0.0) $ & & \\    
                 & $ P = 1.0, \;\; \rho = 1.0$  
                 & $ P = 367.5, \;\; \rho = 25.48$ & & \\
   \hline
    Slow Shock 
                 & $ u^i = (1.53,0.0,0.0)$ 
                 & $ u^i = (.9571,-0.6822,0.00)$  
                 & $ n = 200 $ & $ t = 2.0 $      \\ 
   $(\mu = 0.5)$ 
                 & $ B^i = (10.0,18.28,0.0) $ 
                 & $ B^i = (10.0,14.49,0.0) $ & & \\    
                 & $ P = 10.(!), \;\; \rho = 1.0$  
                 & $ P = 55.36, \;\; \rho = 3.323$ & & \\
   \hline
    Switch-off Fast    
                 & $ u^i = (-2.0,0.0,0.0)$ 
                 & $ u^i = (-0.212,-0.590,0.0)$  
                 & $ n = 150 $ & $ t = 1.0 $      \\ 
    Rarefaction
                 & $ B^i = (2.0,0.0,0.0) $ 
                 & $ B^i = (2.0,4.710,0.0) $ & & \\    
                 & $ P = 1.0, \;\; \rho = 0.1$  
                 & $ P = 10.0, \;\; \rho = 0.562$ & & \\
   \hline
    Switch-on Slow    
                 & $ u^i = (-0.765,-1.386,0.0)$ 
                 & $ u^i = (0.0,0.0,0.0)$  
                 & $ n = 150 $ & $ t = 2.0 $      \\ 
    Rarefaction
                 & $ B^i = (1.0,1.022,0.0) $ 
                 & $ B^i = (1.0,0.0,0.0) $ & & \\    
                 & $ P = 0.1, \;\; \rho = 1.78 \times 10^{-3}$ (!)  
                 & $ P = 1.0, \;\; \rho = 0.01$ & & \\
   \hline
    Alfv\'en wave  
                 & $ u^i = (0.0,0.0,0.0)$ 
                 & $ u^i = (3.70,5.76,0.00)$  
                 & $ n = 200 $ & $ t = 2.0 $      \\ 
   $(\mu = 0.626)$ 
                 & $ B^i = (3.0,3.0,0.0) $ 
                 & $ B^i = (3.0,-6.857,0.0) $ & & \\    
                 & $ P = 1.0, \;\; \rho = 1.0$  
                 & $ P = 1.0, \;\; \rho = 1.0$ & & \\
   \hline
    Compound wave  
                 & $ u^i = (0.0,0.0,0.0)$ 
                 & $ u^i = (3.70,5.76,0.00)$  
                 & $ n = 200 $ & $ t = 0.1,0.75,1.5$      \\ 
                 & $ B^i = (3.0,3.0,0.0) $ 
                 & $ B^i = (3.0,-6.857,0.0) $ & & \\    
                 & $ P = 1.0, \;\; \rho = 1.0$  
                 & $ P = 1.0, \;\; \rho = 1.0$ & & \\
   \hline
    Shock Tube 1    
                 & $ u^i = (0.0,0.0,0.0)$  
                 & $ u^i = (0.0,0.0,0.0)$  
                 & $ n = 400 $ & $ t = 1.0 $      \\ 
                 & $ B^i = (1.0,0.0,0.0) $     
                 & $ B^i = (1.0,0.0,0.0) $ & & \\    
                 & $ P = 1000., \;\; \rho = 1.0 $  
                 & $ P = 1.0, \;\; \rho = 0.1$ (!) & & \\
   \hline
    Shock Tube 2    
                 & $ u^i = (0.0,0.0,0.0)$  
                 & $ u^i = (0.0,0.0,0.0)$  
                 & $ n = 500 $ & $ t = 1.0 $      \\ 
                 & $ B^i = (0.0,20.0,0.0) $     
                 & $ B^i = (0.0,0.0,0.0) $ & & \\    
                 & $ P = 30., \;\; \rho = 1.0 $  
                 & $ P = 1.0, \;\; \rho = 0.1$ (!)& & \\
   \hline
    Collision    
                 & $ u^i = (5.0,0.0,0.0)$  
                 & $ u^i = (-5.0,0.0,0.0)$  
                 & $ n = 200 $ & $ t = 1.22 $      \\ 
                 & $ B^i = (10.0,10.0,0.0) $     
                 & $ B^i = (10.0,-10.0,0.0) $ & & \\    
                 & $ P = 1.0, \;\; \rho = 1.0 $  
                 & $ P = 1.0, \;\; \rho = 1.0$ & & \\
   \hline
   \hline
   \end{tabular} 

   \caption{ Parameters of the test simulations.   $\mu$ is the wave speed
      relative to  the grid.   The second and  the  third columns give the
      left and the right states for an initial discontinuity at the origin
      at  $t=0$.  In  the  case of  Alfv\'en  wave  these are the   states
      connected via the wave. The fourth column  gives the number of cells
      in the  computational  domain. The fifth column   gives the time  at
      which the  results of the simulations  are compared with the exact
      solution.}
   \label{tab-a}
   \end{center}  
\end{table*} 

In addition, the bottom panel of Figure 7 in Komissarov(1999) shows 
$b^y$ but not $B_y$ as it is wrongly indicated in the figure.

\end{document}